\documentclass[12pt]{article}

\usepackage[utf8]{inputenc}
\usepackage[english]{babel}
\usepackage{textcomp}
\usepackage{amsfonts}
\usepackage{amsmath}
\usepackage{amssymb}
\usepackage{mathrsfs}
\usepackage[margin=2.2cm]{geometry}
\usepackage{graphics}
\usepackage{color}
\usepackage{epsfig}
\usepackage{upgreek}

\usepackage{slashed}

\usepackage{sidecap}
\usepackage{caption}
\usepackage{subcaption}

\usepackage[dvipsnames]{xcolor}

\begin{document}

\begin{center}
    {\bf \Large  Dark matter mixing within the seesaw type II mechanism }\\[0.3cm]
    {\bf \Large in the left-right symmetric model\footnote{To appear in     Phys.Part.Nucl.Lett.}}
\end{center}

\begin{center}
{\large M.~Dubinin}$^{*}$, 
{\large E.~Fedotova}$^{*}$,
{\large D.~Kazarkin}$^{*,\dagger}$
\\
\vskip .7cm
{\footnotesize
$^{*}$ Skobeltsyn Institute of Nuclear Physics (SINP MSU), 
M.V. Lomonosov Moscow State University, Leninskie gory, GSP-1, 119991 Moscow, Russia\\[0.3cm]
$^\dagger$ Physics Department, Lomonosov Moscow State University,
Leninskie gory, GSP-1, 119991 Moscow, Russia
}
\end{center}
\vskip 1cm
\begin{abstract}

The seesaw type II mechanism is considered within the framework of a left-right chiral model with a gauge group $SU(2)_L\times SU(2)_R\times U(1)$, the lepton sector of which includes three generations of heavy  Majorana neutrinos. The dependence of the mixing parameters of the lightest sterile neutrino as a dark matter particle on the scales of left-right symmetry breaking in the case of direct and inverse hierarchies of active neutrino masses is analyzed.  

\end{abstract}


\label{sec:intro}
\section*{Introduction}

The most interesting extensions of the Standard Model (SM) are models with a group of gauge symmetry $SU(2)_L \times SU(2)_R \times U(1)_{B-L}$ \cite{history1, history2, history3} (left-right chiral models, hereinafter referred to as LRSM). Within the LRSM, small neutrino masses are generated using the seesaw mechanism \cite{seesaw_general}, and there are various possibilities for selecting a candidate particle for the role of dark matter (DM). In the case of a minimal set of LRSM fields, the lightest sterile neutrino can act as a DM particle, whose mass is ${\cal O}$(keV), the decay time is longer than the lifetime of the Universe, the one-loop decay width to $\gamma \nu$ is marginally small and the velocities at the freeze-out stage are far from relativistic ("warm DM"). 

Like cold DM particles, warm dark matter particles with a mass of about keV are suitable for the formation of large-scale structures in the Universe \cite{warmDM}. If the scale of the left-right symmetry breaking is not significantly different from the electroweak scale, the relic density of right-chiral neutrinos DM, formed due to the oscillation mechanism with standard neutrinos, is approximately the same, however, for large mass scales, too  many right-chiral neutrinos of DM are formed \cite{warmDM}. The problem of overproduction can be circumvented by including the additional contribution of entropy that occurs during the decays of long-lived particles of large mass \cite{injection}. This possibility within the framework of the LRSM was considered in \cite{injection_lr}, where a lower permissible limit for the mass of the right-handed gauge boson $m_{W_R} \sim 10$ TeV, unattainable for the Large Hadron Collider (LHC), was obtained. Note that LRSM in the case of $v_R$ [vacuum expectation value (VEV)] decoupling passes into the scenario of the Neutrino Minimal Standard Model ($\nu$MSM) \cite{nu_msm}. 

Consistent parametric scenarios of $\nu$MSM mixing have been considered in detail in the literature \cite{nu_msm_scenarios}, including explicit diagonalization \cite{casas_ibarra} in the case of three generations of sterile Majorana neutrinos \cite{own_numsm_1, own_numsm_2}. Further refinements of the results \cite{injection_lr} revealed an additional "mass window" $W_R$ in the vicinity of 5 TeV \cite{window_5}, which is an experimentally accessible mass region for collaborations ATLAS and CMS. 

For the scenarios described above, it is of great interest to study the sensitivity of the mixing parameters of active and sterile neutrinos from the VEVs $v_R$ and $v_L$ of the left-right symmetry breaking in the case of warm DM particles at the 10 keV mass scale. If the issue is not very difficult for one generation of sterile neutrinos, the real scenario of three generations of sterile neutrinos considered below is significantly more complicated due to the need to diagonalize the matrix of neutrino states of dimension $6 \times 6$ with the following transformation of the weak current interactions to the mass basis.

\section{The full mixing matrix and the seesaw type II mechanism}

The minimal LRSM scalar sector includes three Higgs multiplets, bidoublet $\phi$ and left and right triplets $\Delta_{L,R}$ with vacuum expectation values and quantum numbers \cite{duka99}
\begin{eqnarray}
    \langle \phi \rangle = \frac{1}{\sqrt{2}}\left( \begin{array}{cc}
    k_1 & 0\\ 0& k_2
    \end{array} \right), &&
    \langle \Delta_{L(R)} \rangle = \frac{1}{\sqrt{2}}\left( \begin{array}{cc}
    0&0 \nonumber 
    \\ 
    v_{L(R)} &0
    \end{array} \right), \quad
    \begin{array}{cccccc}
    \phi: &   (1_C,  2_L,   2_R,  0_{B-L})   \\ 
    \Delta_{L(R)}: &  (1_C, 3(1)_L,   1(3)_R, 2_{B-L}),
    \end{array}
    \label{vev_Delta}
    \nonumber
\end{eqnarray}
where $\sqrt{k_1^2+k_2^2}=246$ GeV. 
Two-stage scheme of symmetry breaking is used, first $SU(2)_R \times U(1)_{B-L}$ to $U(1)_Y$ breaking at the scale $v_R\sim{\cal O}$ (TeV) 
due to the $\Delta_R$ field acquiring a nonzero VEV $v_R$; then breaking of $SU(2)_L\,\times\,U(1)_Y$ to $U(1)_Q$ by acquiring VEV's $\langle\phi\rangle, \langle\Delta_L\rangle$.
The VEV $v_L$ is defined by the condition of minimizing the Higgs potential ('VEV seesaw relation')
$
v_L =v_R^{-1} (\beta_2k_1^2+\beta_1k_1k_2+\beta_3k_2^2)/(2\rho_1-\rho_3),
$
where $\beta_i, \rho_k$ -- couplings in the Higgs potential.
One can see that at $v_R \sim {\cal O}$(TeV) the allowed values of $v_L$ are extremely small or zero in the limiting case.

Neutrino masses appear after the diagonalization of the $6 \times 6$ mass matrix
\begin{equation} 
    M_\nu =
        \left(
            \begin{tabular}{cc}
            $M_L$ & $M_D$ \\
            $M_D^T$ & $M_R$
            \end{tabular} 
        \right) \quad \text{with } \quad 
    M_D = \frac{h_L k_1 + \tilde{h}_L k_2}{\sqrt{2}}, \quad
    M_{L,R} = \sqrt{2} h_M v_{L,R}, \nonumber 
\end{equation}
where $h_L, \tilde{h}_L, h_M$ are real-valued Yukawa interaction matrices. The full $6\times6$ mass matrix in the neutrino sector is a complex-valued symmetric matrix which can be written as $M_\nu=\mathcal{U}\hat{M}_\nu \mathcal{U}^T$ (so-called Takagi factorization), where $M_\nu$ is the mass matrix in the gauge (flavor) basis, $\mathcal{U}$ is a unitary matrix, $\hat{M}_\nu = diag(\hat{m}, \hat{M})$ is a diagonal positively defined matrix, $\hat{m}=diag(m_1,m_2,m_3)$, $\hat{M}=diag(M_1,M_2,M_3)$ are $3\times 3$ matrices of light and heavy neutrinos. 
Introducing the form \cite{nu_msm_scenarios, casas_ibarra}
\begin{equation*}
    \mathcal{U}= \mathcal{W} \left(\begin{array}{cc}
    U_\nu & 0 \\
    0 & U_N^* 
    \end{array}\right),
\qquad \text{ where }
     \mathcal{W} = \exp
    \left( \begin{array}{cc}
    0 & \theta \\
    -\theta^{\dagger} & 0
    \end{array} \right),
\end{equation*}
where $U_\nu, U_N$ are unitary matrices and $\theta < I$, we arrive to the following seesaw type II formula
\begin{equation*}
         m_\nu = M_L - M_D M_N^{-1} M_D^T,
\end{equation*}
and the relation between flavor and mass states of neutrinos can be presented as 
\begin{equation*}
    \nu_L= U_{\rm PMNS} P_L \upnu + \Theta P_L N, \qquad
          \nu_R= -\theta^T U_{\nu}^* P_R \upnu + U_N P_R N,
\end{equation*}
where $P_{L,R}$ is a left- (right-) handed projector, $U_{\rm PMNS} \simeq (1-1/2 \theta \theta^\dagger) U_\nu$, $\Theta \simeq \theta U_N^*$ are mixing matrices. With the help of these matrices, charged and neutral currents of neutrinos can be expressed in the form 
    \begin{align*}
         {\cal L}^\upnu_{NC}  =&  \frac{1}{2}
        \sum_{X=Z_1, Z_2} \overline{\upnu} \gamma^\mu  X_\mu 
        \left[ 
         (U^\dagger U) a_X^L P_L + (U^T \Theta^*) (\Theta^T U^*) a_X^R P_R
         \right] \upnu,
        \\
         {\cal L}^\upnu_{CC} =& \frac{g}{\sqrt{2}}  
        \sum_{X=W_1, W_2}
        \overline{l} \gamma^\mu X_{\mu}^- \left[
        U \, a^L_X P_L +  (\Theta^T U^*) a^R_X P_R
        \right] \upnu 
        +\text{h.c.},
        \\
        {\cal L}^N_{NC}  =&  \frac{1}{2} 
\sum_{X=Z_1, Z_2} \overline{N} \gamma^\mu  X_\mu 
\left[
(\Theta^\dagger \Theta) a_X^L P_L + a_X^R P_R
\right] N \\
&+ 
\frac{1}{2} \left(
\sum_{X=Z_1, Z_2} \overline{\upnu} \gamma^\mu  X_\mu 
\left[
(U^\dagger \Theta) a_X^L P_L - (U^T \Theta^*) a_X^R P_R
\right] N + {\text{h.c.}} \right),\\
{\cal L}^N_{CC} =& \frac{g}{\sqrt{2}} \sum_{X=W_1, W_2}  \overline{l} \gamma^\mu X_{\mu }^- (\Theta a_X^L P_L - a_X^R P_R) N + \text{h.c.},
    \end{align*}
\noindent
where $U=U_{\rm PMNS}$, $a^L_{W_1}=-a^R_{W_2}=\cos \xi$, $a^L_{W_2}= a^R_{W_1}=\sin \xi$, and 
\begin{align*}
& a_{Z_1}^L = e \left[t_W^{-1} c_\phi +t_W (c_\phi -
 c_{2 W}^{-1/2} s_\phi 
 ) \right], \qquad 
a_{Z_1}^R = -e s_\phi \left[
2 c_{2W}^{1/2} s_{2W}^{-1} + t_W c_{2W}^{-1/2}
\right], \\
& a_{Z_2}^L =
e \left[
t_W^{-1} s_\phi + t_W (s_\phi + c_{2W}^{-1/2} c_\phi) 
\right], \qquad
a_{Z_2}^R = e c_\phi \left[
2 c_{2W}^{1/2} s_{2W}^{-1} + t_W c_{2W}^{-1/2}
\right],
\end{align*}
$c_\phi = \cos \phi$, 
$s_{2W} = \sin 2 \theta_W$,
$t_W = \tan \theta_W$, $g=e/s_W$,
$\theta_W$ is the Weinberg angle,
$\phi, \xi$ are mixing angles in the gauge sector
and it is assumed that in the LRSM $V_{L,R}^l = I$, $U_N=I$, $g_L = g_R = g$, where $V_{L,R}$ are unitary matrices in the charged leptons sector, see \cite{duka99}.
The presence of additional gauge bosons modifies neutrino currents compared with the ones in the $\nu$MSM. 
The mixing matrix $\Theta$ of the active neutrino and the sterile neutrino mass state (also known as Heavy Neutral Lepton or HNL) can be parameterized using redefined mass matrix of active neutrinos 
\begin{equation*}
     \Theta = i U_{\rm PMNS} \sqrt{\tilde{m}} \Omega \sqrt{\hat{M}^{-1}}, \quad 
     \text{where} \quad
     \tilde{m} = \hat{m} - U_{\rm PMNS}^{-1} M_L (U_{\rm PMNS}^T)^{-1},
\end{equation*}
where $\Omega $ is an arbitrary complex orthogonal matrix. 
In the approximation $\theta^2 \ll I$, $\hat{m} \ll \hat{M}$ and assuming $U_N=I$ one can use
\begin{equation}
     h_{M} \simeq \frac{1}{\sqrt{2}v_{R}}(\theta^{\dagger}U_{\nu}\hat{m}U_{\nu}^{T}\theta^{*}+U_{N}^{*}\hat{M}U_{N}^{\dagger}) \simeq \frac{\hat{M}}{\sqrt{2}v_{R}}, \nonumber
\end{equation}
therefore,
\begin{equation}
     \tilde{m} \simeq \hat{m}-\frac{v_{L}}{v_{R}}U_{\mathrm{PMNS}}^{\dagger}\hat{M} U_{\mathrm{PMNS}}^*.
     \label{m_tilde}
\end{equation}


\section{Warm dark matter neutrino mixing in LRSM}

The lightest HNL with mass of the order of $1-10~\text{keV}$ is
described by the mixing parameter $\Theta_{\alpha 1}$  and the effective
mass parameter $m_D^{dm}$ introduced in \cite{ own_numsm_1, own_numsm_3}
    \begin{equation*}
    \label{eq:mdm:1}
         M_1 \sum \limits_\alpha |\Theta_{\alpha 1}|^2 \equiv m_D^{dm} = \sum \limits_\alpha |U_{\alpha i} (\sqrt{\tilde{m}})_{ij} \Omega_{j1}|^2 = |\sqrt{\tilde{m}}|^2_{kn} \Omega_{n1} \Omega^*_{k1}. 
    \end{equation*}
Considering the lightest HNL as a DM candidate, the following astrophysical and cosmological constraints are imposed on it
\begin{eqnarray}
    \label{eq:lifetime_sec}
        \tau_{N_1} &\sim &  3\times10^{22}\left(\frac{M_1}{1\mbox{~keV}}\right)^{-4} \Big({\frac{m_D^{dm}}{1\text{~eV}}}\Big)^{-1} \text{sec} ~>~H_0^{-1}\simeq 10^{17}~\text{sec},\\
        \Omega_{N_1}h^2 & \simeq &  \left(\frac{m_D^{dm}}{10^{-5} \text{~eV}} \right) \left( \frac{M_1}{10~\text{keV}} \right) \leq \Omega_{DM}h^2 = 0.12,
    \label{cond:omega}
\end{eqnarray}
where (\ref{eq:lifetime_sec}) is the lifetime limitation for dark matter HNL which is quasi-stable due to small mixing with active neutrinos.
Non-observation of radiative one-loop decay $N_1 \to \gamma, \nu$ with distinct $E_\gamma = M_1/2$ leads to stronger lifetime limit $\tau_{N_1} > 10^{25}$ sec, see \cite{vysotsky,xray}.
The condition (\ref{cond:omega}) follows from the relic density limit of a heavy neutrino $N_1$ due to oscillations between $\nu - N_1$ states (non-resonant overproduction limit).
Summarizing the constraints above, we get 
        \begin{equation*}
        m_D^{dm} < 10^{-5}~\text{eV} \times \min\Big\{(M_1[\text{keV}])^{-1}, 300\times (M_1[\text{keV}])^{-4} \Big\}.
    \end{equation*}
Note that in the limiting case of the $\nu$MSM model $v_L =$ 0, so $\tilde{m} = \hat{m}$.

For $v_L = 0$ these strong constraints assign an explicit form of $\Omega$-matrix for normal/inverse hierarchy, see \cite{own_numsm_3}, for NH $\Omega_{NH}:~\Omega_{j1} \to \delta_{j1}$, for IH $ \Omega_{IH}:~\Omega_{j1} \to \delta_{j3}$, so $m_D^{dm} (v_L =0) = m_{\text{light}}$. We use this form of $\Omega$ as a benchmark scenario for $v_L \neq 0$. Due to the compensating contribution of the VEV of the left Higgs triplet when $m_{\rm light} \sim  \frac{v_L}{v_R} \max\{M_2,M_3\}$ (see \eqref{m_tilde}), the dark matter mixing components $U_1^2 = m_D^{dm}/M_1$ can be reduced without suppressing the mass of the lightest active neutrino.
More accurate calculations require numerical matrix square root methods.  The dips in the curves reflect the values of $v_L$ at which the square roots in the nontrivial matrix $\sqrt{\tilde{m}}$ become zero. The numerical values of $v_L$ are defined by the components of the PMNS matrix. These results are shown in Fig. 1.

\begin{figure}[t]
    \begin{center}
    \begin{minipage}[h]{0.49\linewidth}
        \center{\includegraphics[width=1\linewidth]{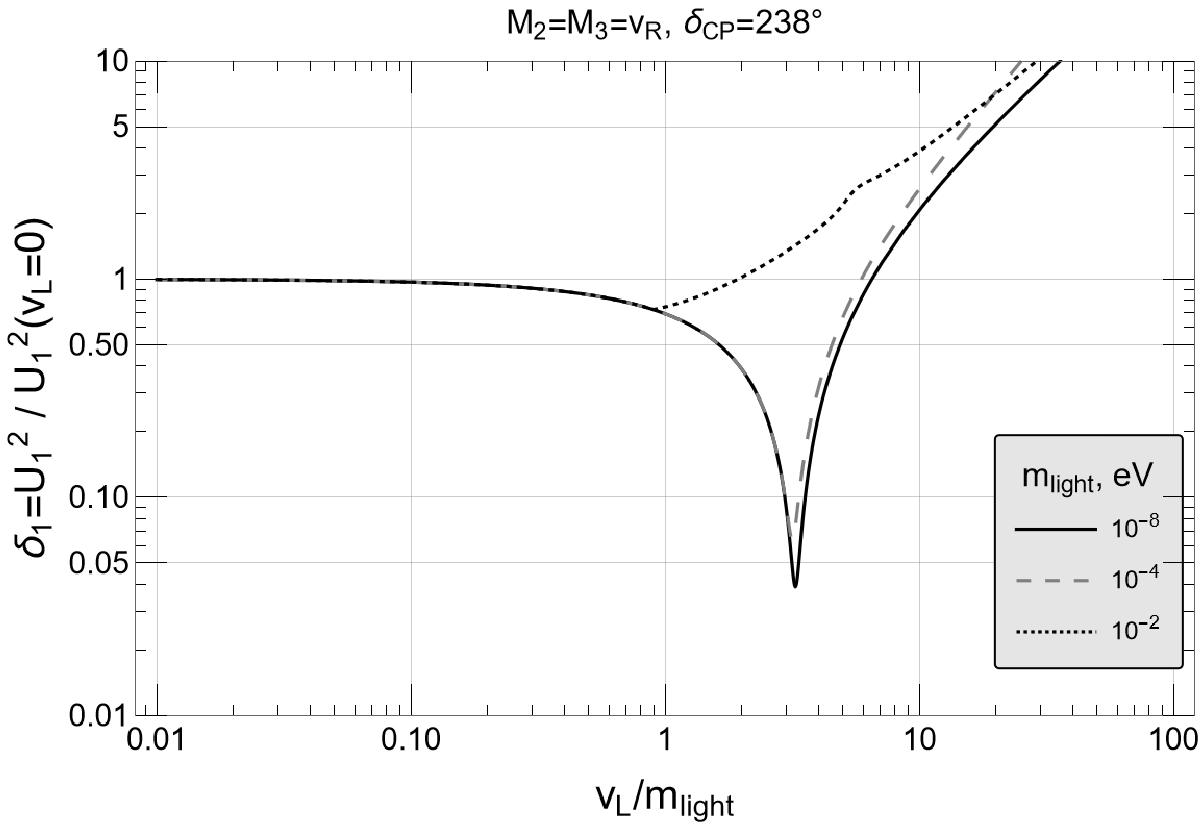}}\\
    \end{minipage}
    \begin{minipage}[h]{0.49\linewidth}
        \center{\includegraphics[width=1\linewidth]{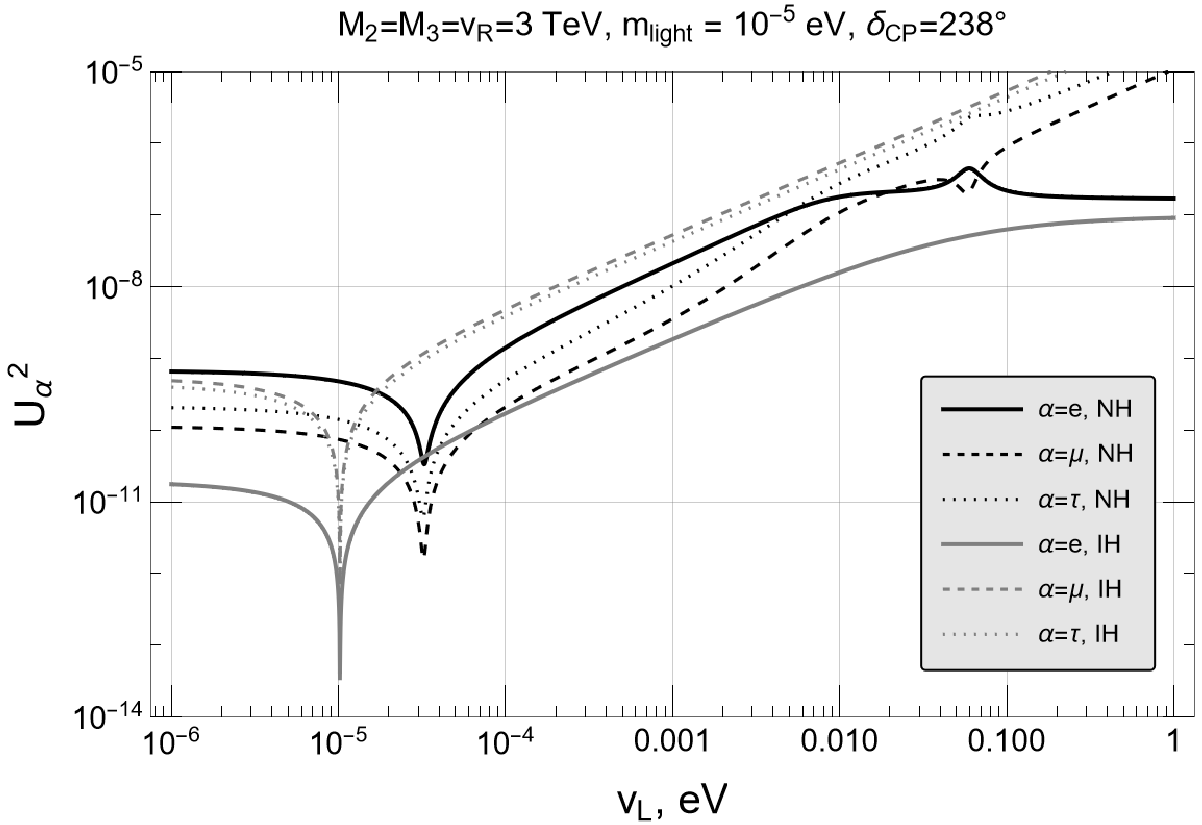}}\\
    \end{minipage}
    \vspace{-3mm}
        \caption{\textbf{Left:} Illustration of $U_1^2$ dark matter mixing parameter decreasing with a fixed $m_{light}$ scale due to the contribution of nonzero VEV of the left Higgs triplet $v_L$. $\delta_1$ is the ratio of mixing parameter $U_I^2$ with nonzero $v_L$ to the same parameter but with $v_L=0$. Here, we consider normal hierarchy of neutrino masses and the set of parameters is selected as follows: $\Omega_{k1} = \delta_{k1}$ ($\nu$MSM-benchmark), $M_2 = M_3 = v_R$. \textbf{Right:} dependence of the mixing components $U_\alpha^2$ on the scale of the left Higgs triplet VEV $v_L$ for both hierarchies and a fixed mass of the lightest active neutrino $m_{light} = 10^{-5}$ eV.
         For both figures, the CP-phase in the PMNS matrix is fixed at $\delta_{CP} = 238^o$.}
        \end{center}
    \vspace{-5mm}
    \label{fig:1}
\end{figure}

\section{Summary}

We considered the lightest sterile neutrino as the dark matter particle in the framework of the LRSM and derived a modified seesaw type II expression for the mixing matrix. 
Main results for the LRSM mixing
$U_{I}^2 = \sum|\Theta_{\alpha I}|^2$
are shown in Fig.1.
If the HNL masses are of the order of $v_R$, then it follows from (\ref{m_tilde}) that $\tilde{m}$ is comparable with $\hat{m}$,
so seesaw type II mass redefinition strongly influences LRSM results
when $v_L \sim m_{\nu_i}$. 
Extremely strong sensitivity of mixing with respect to
$v_L$ scale is observed. At some nonzero $v_L$, extremely small values of mixing parameters are possible, with important consequences for phenomenology.

In the case when $M_{N_i} \ll v_R$, the second term in (\ref{m_tilde}) can always be ignored. 
For example, in the case of $M_{1} \sim {\cal O}$(keV), $M_{2(3)}=m_\pi+m_{\mu(e)}$, 
$v_R \sim \cal O$(TeV)  the additional contribution in  (\ref{m_tilde}) is of $10^{-6} v_L$.
Then the representation for the mixing parameter coincides with the one in the $\nu$MSM.

\label{sec:funding}
\section*{Funding and acknowledgment} 

The research was carried out within the framework of the scientific program of the
National Center for Physics and Mathematics, project ''Particle Physics and Cosmology''.
The work of D.K. was supported by the Theoretical Physics and Mathematics Advancement Foundation ”BASIS” (23-2-2-19-1).
D.K. would like to thank A.Roitgrund
for providing a version of the MLRSM model in the {\it FeynRules} format for testing.

\end{document}